\documentstyle[12pt,epsf,epsfig]{article}
\input{psfig}
\newlength{\dinwidth}
\newlength{\dinmargin}
\newlength{\extraspace}
\newlength{\extraspaces}
\newcommand{\be}{\begin{equation}
\addtolength{\abovedisplayskip}{\extraspaces}
\addtolength{\belowdisplayskip}{\extraspaces}
\addtolength{\abovedisplayshortskip}{\extraspace}
\addtolength{\belowdisplayshortskip}{\extraspace}}
\newcommand{\ee}{\end{equation}}
\newcommand{\bdm}{\begin{displaymath}
\addtolength{\abovedisplayskip}{\extraspaces}
\addtolength{\belowdisplayskip}{\extraspaces}
\addtolength{\abovedisplayshortskip}{\extraspace}
\addtolength{\belowdisplayshortskip}{\extraspace}}
\newcommand{\edm}{\end{displaymath}}
\renewcommand{\thefootnote}{\fnsymbol{footnote}}
\def\simlt{\mathrel{\lower2.5pt\vbox{\lineskip=0pt\baselineskip=0pt
           \hbox{$<$}\hbox{$\sim$}}}}
\def\simgt{\mathrel{\lower2.5pt\vbox{\lineskip=0pt\baselineskip=0pt
           \hbox{$>$}\hbox{$\sim$}}}}

\newcommand{\ls}[1]
   {\dimen0=\fontdimen6\the\font
    \lineskip=#1\dimen0
    \advance\lineskip.5\fontdimen5\the\font
    \advance\lineskip-\dimen0
    \lineskiplimit=.9\lineskip
    \baselineskip=\lineskip
    \advance\baselineskip\dimen0
    \normallineskip\lineskip
    \normallineskiplimit\lineskiplimit
    \normalbaselineskip\baselineskip
    \ignorespaces}

\catcode`@=11
\newcount\@tempcntc
\def\@citex[#1]#2{\if@filesw\immediate\write\@auxout{\string\citation{#2}}\fi
  \@tempcnta\z@\@tempcntb\m@ne\def\@citea{}\@cite{\@for\@citeb:=#2\do
    {\@ifundefined
       {b@\@citeb}{\@citeo\@tempcntb\m@ne\@citea\def\@citea{,}{\bf ?}
\@warning
       {Citation `\@citeb' on page \thepage \space undefined}}%
    {\setbox\z@\hbox{\global\@tempcntc0\csname b@\@citeb\endcsname\relax}%
     \ifnum\@tempcntc=\z@ \@citeo\@tempcntb\m@ne
       \@citea\def\@citea{,}\hbox{\csname b@\@citeb\endcsname}%
     \else
      \advance\@tempcntb\@ne
      \ifnum\@tempcntb=\@tempcntc
      \else\advance\@tempcntb\m@ne\@citeo
      \@tempcnta\@tempcntc\@tempcntb\@tempcntc\fi\fi}}\@citeo}{#1}}
\def\@citeo{\ifnum\@tempcnta>\@tempcntb\else\@citea\def\@citea{,}%
  \ifnum\@tempcnta=\@tempcntb\the\@tempcnta\else
   {\advance\@tempcnta\@ne\ifnum\@tempcnta=\@tempcntb \else 
\def\@citea{--}\fi
    \advance\@tempcnta\m@ne\the\@tempcnta\@citea\the\@tempcntb}\fi\fi}
\catcode`@=12
\newcommand{\SM}{Standard Model}

\newcommand{\la}{\lambda}
\newcommand{\La}{\Lambda}

\newcommand{\cL}{{\cal L}}

\newcommand{\pr}{Phys.\ Rev.\ }
\newcommand{\prp}{Phys.\ Rep.\ }

\newcommand{\np}{Nucl.\ Phys.\ {\bf B}}
\newcommand{\pl}{Phys.\ Lett.\ {\bf B}}

\newcommand{\zp}{Z. Phys.\ {\bf C}}

\begin{document}
\begin{titlepage}
\begin{flushright}
MADPH-97-984\\
TUM-HEP-267/97\\
hep-ph/9702265\\
\end{flushright}
\vspace{14mm}
\begin{center}
\Large{{\bf Model-independent Constraints on Topcolor from $R_b$}}
\end{center}
\vspace{5mm}
\begin{center}
Gustavo Burdman\footnote{e-mail address: burdman@pheno.physics.wisc.edu}\\
*[3.5mm]
{\normalsize\it Department of Physics, University of Wisconsin,}\\ 
{\normalsize
\it Madison, WI 53706, USA}\\ *[3.5mm] and \\*[3.5mm]
Dimitris Kominis\footnote{e-mail address: kominis@physik.tu-muenchen.de}
\\*[3.5mm]
{\normalsize\it Institut f\"ur Theoretische Physik, 
Technische Universit\"at M\"unchen,}\\
{\normalsize\it James-Franck-Stra\ss e, 85748 Garching, Germany }
\end{center}
\vspace{1.00cm}
\thispagestyle{empty}
\begin{abstract}
We identify and compute the main corrections to $R_b$ in Topcolor
theories. They are due to the top-pions, the pseudo-Goldstone bosons 
associated with the chiral symmetry breaking responsible for the 
top quark mass, and amount to $-(1-4)\%$, 
independent of the details of the Topcolor model. 
We show that the cancellation of this effect against the contributions
from Topcolor vector and scalar resonances is unlikely, given the inherently 
distinct character of the two scales governing the potentially 
cancelling terms. The resulting value of $R_b$ leads to severe constraints
on Topcolor theories.
\end{abstract}
\end{titlepage}
\newpage

\renewcommand{\thefootnote}{\arabic{footnote}}
\setcounter{footnote}{0}
\setcounter{page}{2}

\section{Introduction} 
\vspace{-0.2cm}

The large value of the top-quark mass has led to speculation that the
properties of the top quark may be fundamentally different from those of the
other known fermions. In theories of top condensation, for example, 
a new
strong gauge interaction (Topcolor), which couples to the $t$-quark and 
perhaps
to other fermions as well, drives the formation of a
$\langle \bar{t}t \rangle$ condensate and imparts a large dynamical mass
to the top quark. In the process, the chiral symmetries associated with the
strongly interacting fermions are spontaneously broken and at least
three Goldstone bosons emerge. In 
the earliest incarnation of this idea \cite{bhl}, these Goldstone bosons were
identified with the longitudinal components of 
the $W$ and $Z$ 
in an attempt to simultaneously resolve the puzzle of electroweak
symmetry breaking. The Goldstone
boson decay constant $f_{\pi}$ is approximately given in terms of the 
dynamical
top mass $m_t$ and the scale $\La$, at which Topcolor is assumed to
break spontaneously to the \SM\ gauge group, by the Pagels-Stokar formula
\cite{ps} 
\be
f_{\pi}^2 \approx \frac{N_C}{16\pi^2} m_t^2 \ln\frac{\La^2}{m_t^2}
\label{ps}
\ee
where $N_C$ is the number of colors.
If $f_{\pi}$ is to be identified with the weak scale $v_w\approx 175$~GeV,
then 
$\La$ must be of the order of $10^{15}$~GeV. This, in turn, requires an
extreme fine-tuning of the Topcolor gauge coupling in order to achieve a
hierarchy of thirteen orders of magnitude between 
$m_t$ and $\La$.
Alternatively, one may wish to construct top-condensate theories which do not
suffer from such unnatural hierarchies \cite{hill}. This means that the scale
$\La$ 
of the new gauge interactions should be brought down to a few
TeV or so, depending on the amount of fine-tuning one is prepared to
tolerate. From eq. (\ref{ps}) we then deduce that the Goldstone boson decay 
constant is $f_{\pi} \approx 60$~GeV, a value insufficient to account for the
observed $W$ and $Z$ masses. Consequently a separate mechanism, such as
Technicolor (along with Extended Technicolor (ETC)) \cite{tc}, 
must be invoked to supply the remainder of the weak gauge boson masses and
give mass to the lighter fermions. The second major consequence of this
approach is that the Goldstone bosons associated with the formation of the 
$\langle \bar{t}t \rangle$ condensate (``top-pions") are now physical degrees
of freedom\footnote{More precisely, the 
linear combinations of top-pions and techni-pions which couple directly 
to the
weak currents are absorbed via the Higgs mechanism, while the orthogonal 
linear combinations represent genuine physical states.}.

Along with the light fermions, the top quark is expected to acquire itself a
small hard mass of approximately 
 1~GeV or so, from ETC interactions. This breaks the top
chiral symmetries explicitly and turns the top-pions into massive
pseudo-Goldstone bosons. A simple estimate gives \cite{hill} 
\be
M_{\tilde{\pi}}^2 \approx \frac{N_C}{8\pi^2} \frac{m_{ETC}m_t}{f_{\pi}^2}
\La^2
\gamma 
\label{mpi}
\ee
where $\gamma$ is a radiative correction factor which can be rather large
because 
of the strength of Topcolor interactions. For $m_{ETC} \sim 1$~GeV, this
yields $M_{\tilde{\pi}} \sim (100-300)$~GeV.
Note that, because of its dependence on the parameter $m_{ETC}$, the
top-pion mass represents an energy scale which is largely independent from, 
and much lower than the Topcolor scale $\La$.

In this note we wish to point out that the presence of physical top-pions in
the low-energy spectrum is an inevitable feature of any Topcolor model that 
purports to avoid fine-tuning. The effects of top-pions on low-energy
observables are governed by the scale $M_{\tilde{\pi}}$ (not the topcolor 
scale
$\La$), while the 
top-pion couplings to third generation quarks and to gauge bosons are to a
large degree model-independent. Therefore, a study of top-pion effects at low
energies can be used to test the Topcolor idea as a
whole, rather than constrain particular implementations of it.

With these comments in mind we investigate the impact of top-pions on $R_b$,
the ratio of the $Z$ boson decay rate to $b\bar{b}$ pairs to its total 
hadronic
width. This observable is expected to be sensitive to the Topcolor
dynamics because of the non-trivial Topcolor interactions of the left-handed
$b$-quark. We find that the top-pions lead to a substantial 
reduction of $R_b$ relative to the Standard Model prediction. 
While competing
contributions are expected to arise in 
general \cite{hz}, they are governed by the Topcolor scale $\La$ and cannot
compensate fully for the top-pion effects, despite the fact that they are
formally leading in a $1/N_C$ expansion. The crucial observation here 
is that,
in addition to $1/N_C$, the theory possesses another small parameter, namely
$M_{\tilde{\pi}}^2 / \Lambda^2$, which on symmetry grounds must vanish as
$m_{ETC} \rightarrow 0$ independently of the model or the number of colors.
In Sec.~\ref{lowen} we derive the top-pion couplings to quarks 
in the effective Lagrangian approach.  Their effect on $R_b$ is discussed
in Sec.~\ref{rbintc}. Here we also take into account the possible effects of 
Topcolor vector mesons as well as those of additional scalar resonances
present in specific models. We close with some concluding remarks in 
Sec.~\ref{conclu}.

\section{Low-energy Lagrangian}
\label{lowen}
The Topcolor gauge group is assumed to contain an $SU(3)_1\times SU(3)_2$
subgroup, which is spontaneously broken at an energy scale $\La$ down to the
$SU(3)$ of QCD. The doublet $\psi_L = (t_L \; b_L) ^T$
as well as $t_R$ 
couple to $SU(3)_1$, which is the strongest of the two $SU(3)$ groups. 
The representation assignment of the 
remaining fermions is a model-dependent issue. However it should be made 
in such a way as to cancel gauge anomalies and ensure that no 
condensate of known quarks forms other than $\langle \bar{t}t \rangle$.
Additional fermions may be enlisted in order for these requirements to
be fulfilled. In this section we restrict attention only to the 
interactions of
$\psi_L$ and $t_R$ which are common to all models. 

The quantitative study of this theory \cite{hill}
begins with the simplifying assumption
that, at energies below the scale $\La$ the leading effects of the Topcolor
dynamics are adequately described by a contact term
\be
\cL = -\frac{2\pi\kappa}{\La^2} \left( \bar{\psi}_L \gamma_{\mu}
\frac{\la^A}{2}\psi_L  + \bar{t}_R \gamma_{\mu}\frac{\la^A}{2}
t_R\right)^2
\label{curcur}
\ee
which corresponds to the exchange of massive Topcolor gauge bosons. Here
$\kappa$ is essentially the ``fine-structure constant" of $SU(3)_1$ 
and $\la^A$
are the Gell-Mann matrices. Notice that the Lagrangian (\ref{curcur}) is
invariant under an $SU(2)\times U(1)$ chiral symmetry.
By performing a Fierz transformation one can isolate the scalar
channel, i.e. the contact interaction of Lorentz-scalar color-singlet
bilinears, which is relevant for chiral symmetry breaking:
\be
\cL_{{\rm scalar}}=\frac{4\pi\kappa}{\La^2} \bar{\psi}_L t_R \bar{t}_R \psi_L
\label{fourf}
\ee
Calculations with this effective lagrangian are usually carried out in the
so-called Nambu--Jona-Lasinio (NJL) 
approximation \cite{njl}, which consists in resumming
diagrams involving chains of fermion bubbles (but not loops of such
chains). This approximation is justified in the limit where the number of
colors $N_C$ becomes large.

Alternatively, one may introduce collective variables in the form of 
an auxiliary doublet
field $\phi_0$ \cite{eguchi}. The Lagrangian (\ref{fourf}) is equivalent to
\be
\cL_{{\rm scalar}} = -\la \bar{\psi}_L \phi_0 t_R + {\rm h.c.} 
-\La^2 \phi_0^{\dag} \phi_0 
\label{laux}
\ee
where $\la^2 = 4\pi\kappa$.
The renormalization group evolution of this Lagrangian to
low energy scales generates a gauge-invariant kinetic term as well as
self-interactions for $\phi_0$. In terms of the properly renormalized
field $\phi$ the effective Lagrangian reads
\be
\cL_{{\rm eff}}= |D_{\mu}\phi|^2 - \tilde{\la} \bar{\psi}_L \phi t_R + {\rm
h.c.} - V(\phi) 
\label{effl}
\ee
where
\be
D_{\mu} = \partial _{\mu} -igW_{\mu}^a \tau^a - i g'\frac{Y}{2}B_{\mu},
\ee
$\tau^a$ are $SU(2)$ generators and $Y=-1$ is the hypercharge of $\phi$.
The mass-squared term in the potential $V(\phi)$ is negative leading to
the spontaneous breakdown of the chiral symmetries to a $U(1)$ subgroup.

Note that the Lagrangian (\ref{effl}) is largely independent of the 
details of
the above construction. From an effective field theory point of view, the 
expression (\ref{effl}) is the most general Lagrangian describing a
$SU(2)\times U(1) 
\rightarrow U(1) $ symmetry breaking pattern and involving as the sole 
degrees
of freedom (apart from the quarks and gauge bosons) an entire\footnote{The
existence of a full scalar doublet at low energies is suggested by the NJL
approximation and is a consequence of the necessity to keep the theory 
unitary
up to the scale $\La$. If one chose instead to write down a chiral effective
Lagrangian keeping only the fermions and the Goldstone bosons, one would
violate unitarity at a scale $\sim 4\pi f_{\pi} < \La$.}
scalar doublet
$\phi$. The details of the potential $V(\phi)$ are irrelevant to our
discussion, except that it is such as to cause $\phi$ to develop a nonzero
vacuum expectation value, $f_{\pi}$. Since the top quark acquires 
the bulk of its mass through this mechanism, its Yukawa coupling
to the scalar doublet is fixed: $\tilde{\la}= m_t/f_{\pi}$. 
The remaining couplings are determined by gauge 
invariance. As far as  the effective theory is concerned, 
the value of $f_{\pi}$ is a free parameter. In the numerical
estimates of the following sections we take $f_{\pi}=60$~GeV as suggested by
the NJL approximation to the Pagels-Stokar formula (eq.~(\ref{ps})). The soft
dependence on the generally unknown cutoff $\La$ indicates that sharp
deviations from this value are unlikely. 

In terms of the component fields of the scalar doublet, defined by
\be
\phi = \left( \begin{array}{c} f_{\pi}+\frac{1}{\sqrt{2}} (
h + i\tilde{\pi}^0) \\ \tilde{\pi}^- \end{array} \right) ,
\label{comps}
\ee
the Lagrangian (\ref{effl}) is given by
\begin{eqnarray}
\cL_{{\rm eff}} &= & |D_{\mu}\phi|^2 - m_t \bar{t} t - \frac{m_t}{f_{\pi}}
\left[ \frac{1}{\sqrt{2}}\bar{t}_L(h+i\tilde{\pi}^0)t_R + \bar{b}_L
\tilde{\pi}^- t_R + {\rm h.c.} \right] \nonumber \\
& & - \frac{1}{2}M_h^2 \,h^2 + ({\rm scalar} \; {\rm self-interactions})
\end{eqnarray}
where $M_h$ is the mass of the scalar state $h$. 
As explained above, the ETC
interactions explicitly break the chiral symmetries and cause the states
$\tilde{\pi}$ to acquire masses $M_{\tilde{\pi}^{\pm}}, M_{\tilde{\pi}^0}$.
These are not necessarily 
equal in general, because of the lack of an $SU(2)$ custodial
symmetry. Finally, note that the mixing of $\tilde{\pi}^{\pm,0}$ 
with the Goldstone
bosons of electroweak symmetry breaking introduces a factor of $\cos \theta$ 
to
the fermion couplings of the physical top-pions, where $\sin \theta =
f_{\pi}/v_w$, and generates a coupling between the top-pions and the
right-handed $b$-quark.  The latter is proportional to $m_b$ and will be
neglected. Correcting for the former effect and
keeping only the terms which are relevant to the calculation 
of the next section, we finally obtain
\begin{eqnarray}
\cL_{{\rm eff}}&=& |D_{\mu}\phi|^2 - m_t \bar{t} t - \frac{m_t}{f_{\pi}} \cos
\theta \left[
\frac{1}{\sqrt{2}}\bar{t} i\tilde{\pi}^0 \gamma^5 t + \bar{b}_L \tilde{\pi}^-
t_R + \bar{t}_R \tilde{\pi}^+ b_L \right] \nonumber \\
& & - M_{\tilde{\pi}^{\pm}}^2 \tilde{\pi}^+ \tilde{\pi}^- -
\frac{1}{2} M_{\tilde{\pi}^0}^2 \tilde{\pi}^{0\,2} + ...
\label{effl2}
\end{eqnarray}

\section{$R_b$ in Topcolor}
\label{rbintc}
In Fig.~1 we plot the relative correction to $R_b$ due to top-pion
exchange
as a function of the mass of the charged top-pion. The calculation has been
carried out to one loop with the Lagrangian (\ref{effl2}) and follows closely
the corresponding computation in the Two-Higgs-Doublet model 
\cite{hollik}, as
it can be seen in Fig.~2. The contributions of
the neutral states $\tilde{\pi}^0$ and $h$ are suppressed at this level by a
factor of order $(m_b^2/m_t^2)$ and are ignored. 
The arguments of the previous section
imply that the only uncertainty in this 
result resides in the value of 
$f_{\pi}$. 
Of course, due to the strong couplings of the top-pions,
higher-order corrections are expected to be substantial. We therefore regard
our results only as rough estimates.

\begin{figure}
\center
\hspace{0.25cm}
\psfig{figure=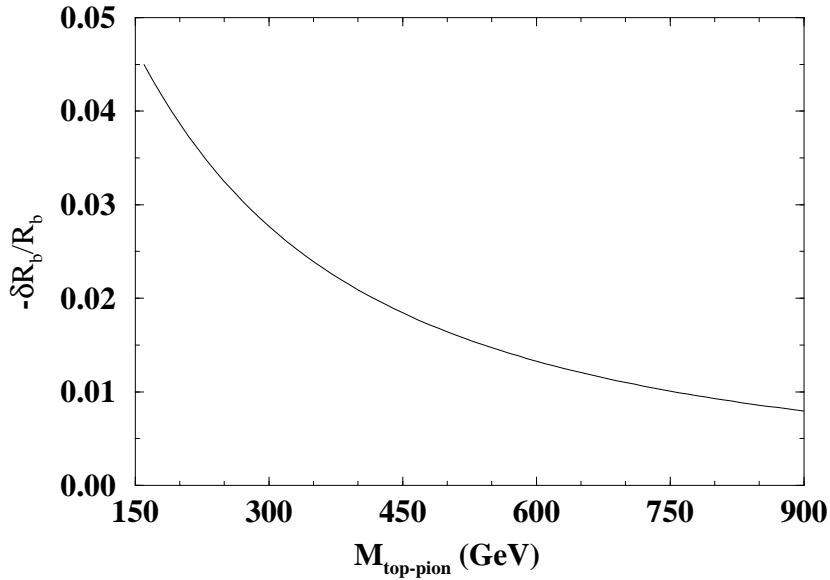,height=3.2in}
\caption{Fractional change in $R_b$ induced by the top-pions of a generic
Topcolor model. Notice that the quantity plotted is the {\em negative} of the
correction to $R_b$.}
\end{figure}

\begin{figure}
\center
\hspace{0.20cm}
\psfig{figure=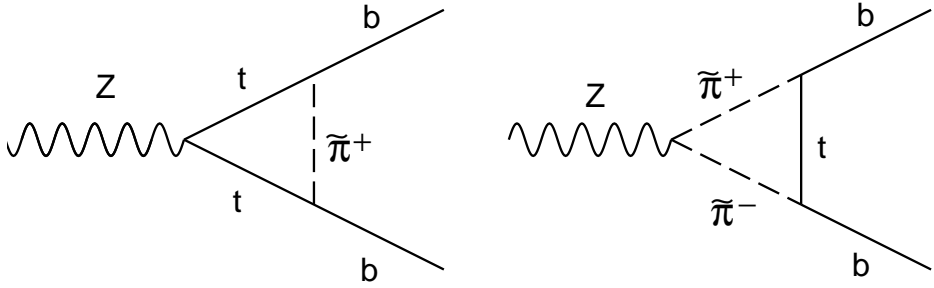,height=1.70in}
\caption{Corrections to the $Z\bar{b}b$ vertex due to top-pions. 
Not shown are the corresponding corrections to the fermion lines.}
\end{figure}

The collaborations at LEP have recently reported a measured value of
$0.2178 \pm 0.0011$~for $R_b$ \cite{warsaw} to be compared with the Standard
Model prediction of 0.2158. 
This allows negative deviations relative to the Standard Model of only 
$0.6\%$ at $3\sigma$, and imposes very
severe constraints on Topcolor models, unless there exist further
contributions to $R_b$,
which can naturally neutralize the top-pion effects.
Candidates for such contributions include the exchange of Topcolor vector
mesons or of additional light scalars which sometimes occur 
in specific models.
We now turn to the study of these states and of their influence on $R_b$,
adopting the NJL approximation as our calculational framework.

As mentioned in Section~2, a Topcolor model is specified when the 
anomaly-free assignment of all fermions is chosen so as to ensure that 
no  SM fermion other than the top quark acquires a large
dynamical mass. In particular, the $b$ quark will get a large mass
if $b_R$ couples to the Topcolor interaction in the same way as $t_R$
does. This need for explicit  
isospin breaking has been addressed in various ways, all of which result
in the presence of additional scalars in the spectrum. Here we show
the effects of this model-dependent aspect of Topcolor on 
electroweak observables in the two scenarios described in \cite{bbhk}. 

First, let us consider the case where the condensation is 
tilted in the top direction by the presence of an extra $U(1)$ interaction
embedded in the weak hypercharge group such that, at the scale 
$\Lambda$, $U(1)_1\times U(1)_2\rightarrow U(1)_Y$. Here the $U(1)_1$
is strongly coupled to the third generation fermions.  The fermions have 
$U(1)_1\times U(1)_2$ assignments such that
the couplings of the remnant $Z'$ to fermions
are analogous to those of the $Z$. Thus the effect of the $Z'$ in the 
top channel adds to that of the Topcolor interaction, whereas it is 
repulsive in the $b$ quark channel. This defines a region in the 
coupling parameter space where only
the coupling in the top channel is super-critical, leading to the breaking
of the top chiral symmetry and to the scalar doublet in (\ref{comps}). 
In addition, in this region the effective interaction in the $b$ channel,
although sub-critical, is strong enough to form rather deeply
bound scalar states \cite{dk}. These make up the doublet
\begin{equation}
\phi_b=\left(\begin{array}{c}
H^+ \\
\frac{1}{\sqrt{2}}\left(H^0+iA^0\right) 
\end{array}\right)
\label{phi_2}
\end{equation}
Their masses can be estimated within the NJL
approximation to be \cite{dk} 
\begin{equation}
M_{H^0,A}\simeq (150-350)~{\rm GeV}, \label{bpi_0} 
\end{equation}
whereas the mass of the charged states is determined by the 
relation 
\begin{equation}
M_{H^\pm}^2=M_{H^0,A}^2 + 2\,m_t^2 . \label{bpi_pm}
\end{equation} 
The presence of this doublet adds the term
\begin{equation}
-\tilde{\lambda} \bar{\psi}_L \phi_b b_R + {\rm h.c.} 
\label{leffb}
\end{equation}
to the effective Lagrangian (\ref{effl}), where as before $\tilde{\lambda} =
m_t/f_{\pi}$. Therefore the charged
scalars $H^\pm$ couple to $b_R$ with strength $\tilde{\lambda}$,
whereas the neutrals $H^0,A^0$ have the same strong coupling to 
$\bar{b}_L b_R$. This results in additional contributions to $R_b$.      
The overall effect of the ``$b$-pions'' $H^\pm$ and $H^0,A^0$ is to 
further decrease $R_b$ by a small amount, making the deviation from 
the experimental result even larger. For instance, for 
$M_{H^0}=350$~GeV and $M_{\tilde{\pi}}=300$~GeV, the value of $R_b$
is reduced by about $3.3\%$. In the absence of the $b$-pions the effect
is $\simeq 2.8\%$. In these estimates we used the relation (\ref{bpi_pm}) for
the $b$-pion masses, which however does not have to be taken literally. It is
well-known \cite{montreal} that the charged scalars in $\phi_b$ tend 
to reduce
$R_b$ while the contribution from the neutrals is positive. The latter 
can only
become dominant, however, if $M_{H^0, A} \ll M_{H^{\pm}}$, in drastic 
departure
from the NJL estimate (\ref{bpi_pm}). To reverse the top-pion effect one 
would
further need $M_{H^0, A} \ll M_{\tilde{\pi}}$, which is counterintuitive.
The relatively mild effect of the ``$b$-pions'' can be traced to the
fact that the $Z$ boson decays predominantly into left-handed $b$-quarks, 
while
in the limit of negligible $m_b$ the scalars
$H^{\pm}$ couple exclusively to $b_R$.

One can further study the effects of vector and axial-vector mesons in
Topcolor by essentially repeating the steps outlined at the beginning of
Section~\ref{lowen} for the scalar doublet $\phi$. The methods employed have
been 
extensively discussed in different contexts by various authors 
\cite{bijnens}.
A vector or axial-vector resonance composed of $b$-quarks would
contribute to $R_b$ by virtue of its mixing with the $Z$. In fact, 
taking into
account this mixing amounts to summing all leading $1/N_C$ contributions
to the $Zb\bar{b}$ vertex. However, an estimate
in the NJL approximation shows that these resonances are very loosely bound,
that is, they have a mass of the order of the Topcolor scale $\Lambda$, and
consequently the resummation hardly improves on the one-loop result of
Ref.~\cite{hz}, according to which $R_b$ receives a positive correction of 
$\delta R_b /R_b \approx 2.5\times 10^{-3}$  for $\La =
2$~TeV falling to $\delta R_b /R_b \approx 5.8\times 10^{-4}$ for $\La =
5$~TeV. This correction cannot compensate for the top-pion 
effects (cf. Fig.~1)
which are governed by 
the -- presumably much lower -- scale $M_{\tilde{\pi}}$.

Another 
way to avoid the formation of a 
$\langle\bar b_L b_R\rangle$ condensate is to have the 
$b_R$ transforming as a $SU(3)_1$ singlet, so that it does not feel
the Topcolor interaction. 
Following ref.~\cite{bbhk}, we notice that the cancelation of anomalies
requires the introduction of a new set of fermion fields, $Q^{a}_{L,R}$,
with $a=1,..., N_Q$. There is a new interaction, $SU(N_Q)$. 
For instance for  $N_Q=3$ 
the following assignments under $SU(3)_Q\times~SU(3)_1\times~SU(3)_2$  
are anomaly-free:
\begin{eqnarray}
(t,b)_L~ (c,s)_L & \simeq& (1,3,1) \nonumber\\ 
t_R &\simeq & (1,3,1) \nonumber\\
Q_R &\simeq & (3,3,1) \nonumber\\
(u,d)_L &\simeq & (1,1,3) \nonumber\\
(u,d)_R ~(c,s)_R &\simeq & (1,1,3) \nonumber\\
b_R &\simeq & (1,1,3)\nonumber\\
Q_L &\simeq & (3,1,3). \nonumber
\end{eqnarray}
The fact that  $(c,s)_L$ transforms 
as a triplet under $SU(3)_1$, in addition to $(t,b)_L$, implies 
that the chiral symmetry is now $SU(4)_L\times U(1)$. The breaking of
this global symmetry by the Topcolor interactions now leads not only
to the scalar doublet $\phi$ defined in Section~2 but also to the
presence of an extra  doublet of Goldstone bosons, the ``charm-top-pions'':
\begin{equation}
\cal C = \left(\begin{array}{c}
\frac{1}{\sqrt{2}}(h_c+i\pi_c) \\
\pi_c^- 
\end{array}
\right) 
\label{c_pions}
\end{equation}
Although in principle massless, they acquire a mass generated by the 
same explicit chiral symmetry breaking responsible for $M_{\tilde\pi}$
and thus of similar value. Their coupling to the right-handed top quark
is equal to that of the top-pions:
\begin{equation}
-\tilde{\lambda}~(\bar{c}~\bar{s})_L~{\cal{C}}~t_R  + {\rm h.c.}
\label{ctp_toq}
\end{equation}
This Topcolor model not only suffers from large deviations
from the measured value of $R_b$, 
but in addition, the couplings in (\ref{ctp_toq}) imply large 
corrections to $R_c$. In Fig.~3 we plot the relative corrections to $R_b$ 
and to $R_c$ 
in this model as a function
of the pseudo-Goldstone boson mass. Also shown is the $3\sigma$ LEP result.
\begin{figure}
\center
\vspace{-1.2cm}\hspace{0.25cm}
\psfig{figure=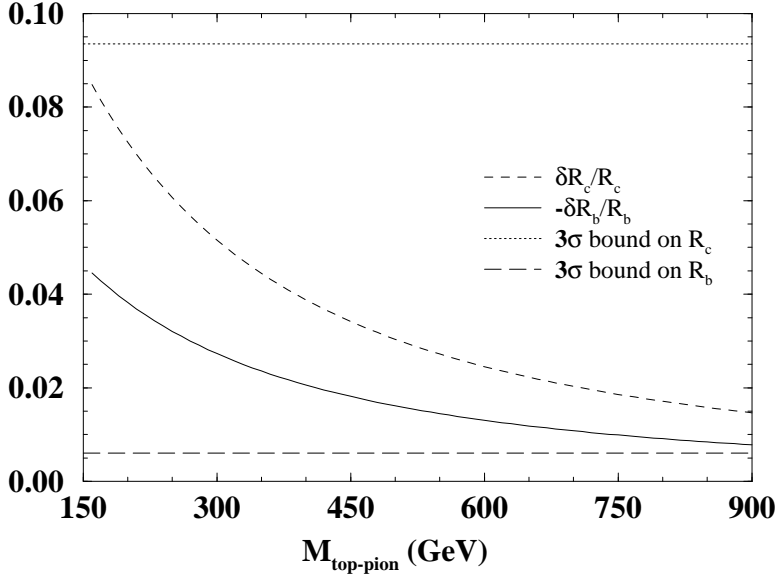,height=3.2in}
\caption{Fractional change in $R_b$ and $R_c$ in the Topcolor model with
strongly coupled $(c,s)_L$ relative to the Standard Model value. The horizontal
lines represent the experimental 3$\sigma$ upper bounds to these
corrections. Note that, as in Fig.~1, the $R_b$ correction is negative.} 
\end{figure}
Although the determination of 
$R_c$ is 
more uncertain, it can be seen that it can potentially
pose an important additional
constraint on these type of models. Note finally that the
decay width of the $Z$ to light hadrons is also affected because 
$R_s$ is reduced by the same amount as $R_b$.
In sum, this particular Topcolor model presents a lot of difficulties
when confronted with observables at the $Z$ pole. 
In general, it is likely that other specific models will give rise to
large deviations not only in $R_b$ but also in $R_c$ and
the $Z$ hadronic width due to the ubiquitous presence of additional scalars
in the spectrum. 

\section{Conclusions}
\label{conclu}

We have seen that it is possible to constrain top condensation theories in
general and Topcolor models in particular by concentrating
on the effects due to the presence of 
top-pions in the spectrum. Not only are these the minimal 
scalar content of any Topcolor model, but they also have a largely 
model-independent coupling to quarks, $\tilde\lambda = m_t/f_{\pi}$,
allowing for  rather general predictions. 
As it can be seen from (\ref{mpi}), 
their masses are expected to be well below the Topcolor scale $\Lambda$. 
This, together with the fact that their decay constant is expected 
to lie in the range $f_{\pi}=(50-70)$~GeV in order to avoid fine-tuning, 
brings about the potential for strong effects, governed by the scale 
$M_{\tilde{\pi}}$ of a few hundred GeV rather than by the Topcolor scale
itself.  
We computed the effect of top-pions in $R_b$ and found it to be a large 
and negative shift with respect to the SM, as shown in Fig.~1. For typical
values of $M_{\tilde{\pi}}$ this 
result is particularly constraining when compared to the
experimental measurements at LEP \cite{warsaw},
which at present point to a positive shift.

The effects of other Topcolor states were taken into account 
in order to investigate the 
possibility of cancellations. As noted in ref. \cite{hz}, the contribution
of the heavy topcolor gauge bosons, or ``top-gluons'', 
to $R_b$ is positive and
small (less then 1\% for realistic Topcolor scales). 
We considered the positive contributions of vector and 
axial-vector states and found that they are
not enough to compensate for the negative shift due to the top-pions. 
We remark that, although the 
precise quantitative results depend on the approximation used, the  
contributions of these states are controlled by the scale $\Lambda$.
Thus, a cancellation as large as the one required seems unlikely given
the two different scales entering. 

Attempts to involve some of the other standard quarks in the strong Topcolor
interactions typically lead to additional light scalars which exacerbate the
effect. 
We calculated the contributions to $R_b$ from these states
in two specific realizations. 
In the model with an extra $U(1)$ we observe that the shift in $R_b$
is slightly more negative than in the sole presence of top-pions. 
The other model considered requires $(c~s)_L$ to be a triplet under the 
strong $SU(3)_1$, i.e. to couple to the top-gluon. 
Here, in addition to the top-pions which modify the $Zb\bar{b}$ vertex in the
same way as in the previous model, there is a new set of pseudo-Goldstone
bosons, the charm-top-pions,  
which induce a very large positive shift to $R_c$. This also conflicts with 
experiment, but the discrepancy is milder than in the case of $R_b$. The
situation is summarized in Fig.~3. There is also a
large negative shift in $R_s$, of the same order as the one in $R_b$. 

We conclude that a large and negative correction to $R_b$ is present in 
Topcolor models with minimal fermionic structure, and that there are no
natural sources of cancellation, at least in the cases examined. 
 
Our quantitative conclusions are of course valid only to the extent that the
approximations we use are justified. Were $f_{\pi}$, for instance, allowed to
float upwards by a factor of two relative to the Pagels-Stokar estimate, the
correction to $R_b$ would be about 6.7 times smaller than the one 
presented in
Fig.~1. Similarly, larger values of $M_{\tilde{\pi}}$ cannot be 
excluded given
the uncertainty in the radiative correction factor $\gamma$ in
eq.~(\ref{mpi}). The top-pion mass can also be enhanced by allowing larger
values of $m_{ETC}$, but this option is rather unattractive.

A possible way out could in principle be provided by an extended fermionic
content of the theory. If the top quark mixes with a new strongly interacting
fermion of electric charge 2/3, but possibly of ``exotic" electroweak quantum
numbers, then the correction to $R_b$ could be naturally smaller than in the
case of no mixing. The challenge is to achieve this while avoiding the
proliferation of pseudo-Goldstone bosons which will emerge from the
breaking of
a necessarily larger chiral symmetry group. Although the simplest models with
additional fermions do not give rise to the desired symmetry breaking 
pattern,
this remains perhaps an avenue to pursue further.

It should be remarked that other electroweak observables
are also affected by the presence of light strongly interacting
top-pions. For instance, 
the $\rho$ parameter receives a two-loop contribution from the latter
in addition to the one from top-gluon exchange considered in ref.~\cite{cdt}.
Any proposed mechanism introducing delicate cancellations in order
to bring $R_b$ in line with experiment should also have a
counterpart in 
the case of all other observables which are sensitive to the presence of
pseudo-Goldstone bosons.

\vspace{0.7cm}
\noindent
{\bf Acknowledgements}

We thank Chris Hill for helpful discussions and a careful reading 
of the manuscript. 
DK thanks the Fermilab Theory Group for their warm hospitality
during the initial stages of this work. His work was supported in
part by the German DFG under contract number Li519/2-1.
The work of GB was supported  by the U.S.~Department of Energy under  
Grant No.~DE-FG02-95ER40896 and the University of 
Wisconsin Research Committee with funds granted by the Wisconsin 
Alumni Research Foundation.


\begin{thebibliography}{99}
\bibitem{bhl} Y.~Nambu, in {\it New Theories in Physics}, Proceedings of
the XI International Symposium on Elementary Particle Physics, Kazimierz,
Poland, 1988, edited by Z.~Adjuk, S.~Pokorski and A.~Trautmann (World
Scientific, Singapore, 1989); Enrico Fermi Institute Report EFI~89-08
(unpublished);
V.~A.~Miransky, M.~Tanabashi and K.~Yamawaki, Phys.~Lett.~{\bf
221B}, (1989) 177; Mod.~Phys.~Lett.~{\bf A4} (1989) 1043;
W.~A.~Bardeen, C.~T.~Hill and M.~Lindner, Phys.~Rev.~{\bf D41},
(1990) 1647.
\bibitem{ps} H.~Pagels and S.~Stokar, \pr {\bf D20} (1979) 2947.
\bibitem{hill} C.~T.~Hill, \pl {\bf 345} (1995) 483.
\bibitem{tc} S. Weinberg, \pr {\bf D19} (1979) 1277;\\
 L. Susskind, \pr {\bf D20} (1979) 2619;\\
 S. Dimopoulos and L. Susskind, \np {\bf 155} (1979) 237;\\
 E. Eichten and K. Lane, \pl {\bf 90} (1980) 125.
\bibitem{hz} C.~T.~Hill and X.~Zhang, \pr {\bf D51} (1995) 3563.
\bibitem{njl} Y.~Nambu and G.~Jona-Lasinio, \pr {\bf 122} (1961) 345; 
{\bf 124} (1961) 246.
\bibitem{eguchi} T. Eguchi, \pr {\bf D14} (1976) 2755.
\bibitem{hollik} A.~Denner, R.~Guth, W.~Hollik and J.~H.~K\"uhn, \zp {\bf 51}
(1991) 695.
\bibitem{warsaw} A.~Blondel, Plenary talk at the 28th International Conf. on
High Energy Physics, Warsaw, July 1996.
\bibitem{bbhk} G.~Buchalla, G.~Burdman, C.~T.~Hill and D.~Kominis, \pr {\bf
D53} (1996) 5185.
\bibitem{dk} D.~Kominis, \pl {\bf 358} (1995) 312.
\bibitem{montreal} P.~Bamert et al., \pr {\bf D54} (1996) 4275.
\bibitem{bijnens} V.~Bernard and U.-G.~Meissner, Nucl.\ Phys.\ {\bf A489}
(1988) 647;
J.~Bijnens, C.~Bruno and E.~de Rafael, \np {\bf390} (1993)
501; J.~Bijnens, E.~de Rafael and H.~Zheng, \zp {\bf62} (1994) 437;
J.~Bijnens,
\prp {\bf 265} (1996) 369.
\bibitem{cdt} R.~S.~Chivukula, B.~A.~Dobrescu and J.~Terning, \pl {\bf 353}
(1995) 289.

\end{thebibliography}
\end{document}